\documentclass[twocolumn, showpacs,preprintnumbers,amsmath,amssymb,pra]{revtex4}
\usepackage[ansinew]{inputenc}
\usepackage{graphicx}
\usepackage{pst-all}
\usepackage{color}
\usepackage{dcolumn}
\usepackage{amsmath}
\usepackage{amsthm}
\usepackage{bm}
\usepackage{layout}
\usepackage{float}
\usepackage{txfonts}
\usepackage{amsfonts}
\usepackage{amssymb}%
\usepackage[ansinew]{inputenc}
\usepackage{graphicx}
\usepackage{amsmath}
\usepackage{amsthm}
\usepackage{bm}
\usepackage{layout}
\usepackage{float}
\usepackage{amsfonts}
\usepackage{amssymb}
\usepackage{array}
\usepackage{multirow,bigdelim}
\usepackage{slashbox}
\usepackage{enumitem}
\usepackage{graphicx}
\usepackage{dcolumn}
\usepackage{bm}
\usepackage{amssymb}
\usepackage{amsfonts}
\usepackage{graphicx}
\usepackage{dcolumn}
\usepackage{bm}

\usepackage{times,epsfig,amssymb,amsmath}

\usepackage{bm}
\usepackage{times,epsfig,amssymb,amsmath}
\usepackage{graphicx}

\begin{document}

\title{Genuine Correlations of Tripartite System}
\author{Liming Zhao,$^{1,2}$Xueyuan Hu,$^2$\footnote{xyhu@iphy.ac.cn}
R.-H. Yue,$^1$ and Heng Fan$^2$\footnote{hfan@iphy.ac.cn}}%
\affiliation{%
$^1$Faculty of Science, Ningbo University, Ningbo 315211, China\\
$^2$Beijing National Laboratory for Condensed Matter Physics,
Institute of Physics,
Chinese Academy of Sciences,
Beijing 100190, China}

\date{\today}

\begin{abstract}
We define genuine total, classical and quantum
correlations in tripartite systems. The measure we propose is based
on the idea that genuine tripartite correlation exists if and only if the
correlation between any bipartition does not vanish. We find in a
symmetrical tripartite state, for total correlation and classical
correlation, the genuine tripartite correlations are no
less than pair-wise correlations. However, the genuine quantum tripartite
correlation can be surpassed by the
pair-wise quantum correlations. Analytical expressions for genuine
tripartite correlations are obtained for pure states and rank-2
symmetrical states. The genuine correlations in both entangled and
separable states are calculated.
\end{abstract}


\maketitle

\section{I. INTRODUCTION}

It is believed that various quantum correlations are
resources in quantum information processing. Such quantum correlations
include entanglement and quantum discord, they are closely related but
different from each other. The study of entanglement
has a long history \cite{entanglementreview} but continuously with
some new applications \cite{cuinaturecomm}.
Quantum discord, introduced in Refs. \cite{Ollivier:2001,vedral1,vedral2}, also has
recently drawn much attention. There is nonzero quantum discord
in general separable states in comparison the entanglement vanishes
in this case. Additionally, entanglement may disappear at a finite time \cite{Yu:2004},
while quantum discord decays asymptotically even at a finite
temperature, which is known as the robustness of quantum discord
\cite{Werlang:2009}. In general it seems that quantum discord is more common
than entanglement in quantum states. For specific applications of
quantum discord in quantum information protocols,
the deterministic quantum computation with one qubit (DQC1) \cite{Datta:2008} and state discrimination
\cite{Roa:2011,Bo:2012} can be realized using states with zero
entanglement but positive quantum correlation. The importance of
quantum discord
is also presented in quantum communication \cite{Ye:2010}. 
Generally, numerical methods is required for calculating quantum
discord. By far, analytical evaluation of quantum discord are
developed for some special two-qubit states, e. g. X states
\cite{luo:2008,Ali:2010}. A simple
method is pointed out to get the analytical quantum discord of a rank-two bipartite
state by relationship with entanglement of formation \cite{Koashi:2004}.

Despite the great potential advantages of multipartite quantum
correlations, the problem of quantifying and characterizing the
correlation contained in the multipartite quantum systems is still
open. Recently, much effort has been devoted in measuring the
quantum correlations in multipartite systems. A global measure is proposed for quantum correlations of multipartite
systems \cite{Rulli:2011}. The quantum dissension which measures the quantum
correlations in a tripartite state is introduced in Ref.
\cite{Agrawal:2011}. Moreover in Ref. \cite{Giorgi:2011},
the genuine total, quantum and classical correlations in multipartite
system are defined by employing relative entropy as a distance measure of
correlations \cite{Modi:2010}. Some general criteria about
the definition of genuine $n$-partite correlations has been given
by Bennett \emph{et al.} in Ref. \cite{Bennett:2011}.

In this paper, we try to provide some new definitions of genuine tripartite
correlations. The genuine total or quantum correlation goes to zero
if and only if there is a bipartition of the tripartite system such
that no total or quantum correlation exist between the two parts. We
obtain the analytical expressions for the genuine
tripartite correlations in pure and rank-two states. Genuine
total, quantum and classical correlations are compared with their
pairwise counterparts. We find that the genuine total and classical
correlations are greater than their pairwise counterparts in symmetrical states, but this
quantitative relation does not hold for quantum correlations. Our
definition of genuine tripartite quantum correlation is no more than
the one defined in \cite{Giorgi:2011}, but the two definitions
coincide for pure states.

The remainder of this paper is arranged as follows. In Sec II, we give our
new definitions of genuine tripartite correlations and investigate
the case of pure and symmetrical tripartite states. In Sec III, we will
illustrate by examples the tripartite genuine quantum discord and
classical correlations in symmetrical systems. Sec IV is  devoted to
the conclusion.

\section{AN OPTIMAL GENUINE TRIPARTITE CORRELATIONS}

 Let us briefly recall the definition of two-particle quantum discord
\cite{Ollivier:2001}:
\begin{equation}
D(\rho_{a:b})\equiv I(\rho_{a,b})-J(\rho_{a:b}),
\end{equation}
where the mutual information
$I(\rho_{a,b})=S(\rho_{a})+S(\rho_{b})-S(\rho_{ab})$, with
$S(\rho)=-tr(\rho\log_2\rho)$, characterizes the total correlations,
and $J(\rho_{a:b})=S(\rho_{a})-\min_{\{E^b_l\}}
S(\rho_{a|{\{E^b_l}\}})$ is called the classical correlation. Here
$\{E^b_l\}$ is positive operator valued measures (POVM) performed on
system $b$ and the conditional entropy is
$S(\rho_{a|{\{E^b_l}\}})=\sum_l p_lS(\rho_{a|{E^b_l}})$, where
$\rho_{a|{E^b_l}}=Tr_b(E^b_l\rho_{ab})/p_l$  and
$p_l=Tr_{ab}(E^b_l\rho_{ab})$.

A state of $n$ particles is said to possess genuine $n$-partite
correlations when it is nonproduct in every bipartite cut, according
to \cite{Bennett:2011}. From this point of view, we can define
genuine tripartite correlations in tripartite states
$\rho_{abc}\equiv\rho$ as
\begin{equation}
T^{(3)}(\rho)\equiv\min[I(\rho_{a,bc}),I(\rho_{b,ac}),I(\rho_{c,ab})],
\end{equation}
where $I(\rho_{i,jk})=S(\rho_{i})+S(\rho_{jk})-S(\rho)$ is the
mutual information between one-qubit part and the left two-qubit
part, which goes to zero if and only if $\rho=\rho_i\otimes
\rho_{jk}$. This definition of genuine total correlation coincides
with the one defined in Ref. \cite{Giorgi:2011}.

Consistently with the definition of $T^{(3)}$, a tripartite state
does not have genuine tripartite quantum discord when there exist a
bipartition such that the quantum correlation between the two parts
is zero. Obviously, $D(\rho_{i:jk})$ equals zero if $D(\rho_{ij:k})$
takes zero, but the converse is not true. Therefore, we define
genuine tripartite quantum discord as:
\begin{equation}
D^{(3)}(\rho)=\min[D(\rho_{a:bc}),D(\rho_{b:ac}),D(\rho_{c:ab})],
\end{equation}
where $D(\rho_{i:jk})=S(\rho_{jk})+S(\rho_{i|jk})-S(\rho)$,
$S(\rho_{k|ij})=\min_{\{E^{ij}_m\}}[S(\rho_{k|\{E^{ij}_m\}})]$, and
$\{E^{ij}_m\}$ is a two-particle POVM operating on $i$ and $j$.
Then, we define  genuine tripartite classical correlations as:
\begin{equation}
J^{(3)}(\rho)=T^{(3)}(\rho)-D^{(3)}(\rho).
\end{equation}

Without loss of generality, we make the following assumption:
\begin{equation}
I(\rho_{c,ab})\leq I(\rho_{a,bc})\leq I(\rho_{b,ac})\label{eq1}.
\end{equation}
The left discussion in this section is based on it. For pure states,
the relative entropy $S(\rho_{c|ab})=0$. By assumption (\ref{eq1}),
we have $S(\rho_{c})=S(\rho_{ab})\leq S(\rho_{a})=S(\rho_{bc}) \leq
S(\rho_{b})=S(\rho_{ac})$. Therefore, we can obtain the genuine
tripartite total, classical correlations and quantum discord:
\begin{eqnarray}
T^{(3)}(\rho)&=&2S(\rho_{c})=2S(\rho_{ab}),\nonumber\\
D^{(3)}(\rho)&=&J^{(3)}(\rho)=S(\rho_{ab})=S(\rho_{c}).
\end{eqnarray}
It means that genuine tripartite classical correlations and quantum
discord are both equal to half of the genuine tripartite total
correlations in a pure state, which coincide with the genuine
correlations defined in Ref \cite{Giorgi:2011}.

When the tripartite quantum system is symmetrical, i. e., the state
of the whole system is invariant under the permutations of the three
parties, the genuine tripartite total, classical and quantum
correlations can be regard as:
 \begin{eqnarray}
T^{(3)}(\rho)&=&S(\rho_{c})+S(\rho_{ab})-S(\rho),\nonumber\\
J^{(3)}(\rho)&=&S(\rho_{c})-S(\rho_{c|ab}),\nonumber\\
D^{(3)}(\rho)&=&S(\rho_{ab})+S(\rho_{c|ab})-S(\rho).
\end{eqnarray}
We derive some properties of the genuine correlations.

$Theorem$. For a symmetrical tripartite quantum state, the genuine
tripartite total and classical correlations is no less than the any
pairwise counterpart, respectively
\begin{eqnarray}
T^{(3)}(\rho)&\geq& T^{(2)}(\rho),\nonumber\\
J^{(3)}(\rho)&\geq& J^{(2)}(\rho),
\end{eqnarray}
where the $T^{(2)}$ and $J^{(2)}$ are pairwise total and classical
correlations respectively.

$Proof$ $of$ $Theory$ --The mutual information does not increase
when discard quantum subsystem: $I(\rho_{c,a}) \leq I(\rho_{c,ab})$
\cite{Nielsen:2000}, it is obvious that genuine tripartite
correlations is no less then pairwise correlations of symmetrical
tripartite systems. For classical correlations, we have
$J^{(3)}(\rho)=J(\rho_{c:ab})$ and $J^{(2)}(\rho)=J(\rho_{c:a})$.
Direct calculations lead to
$J^{(3)}(\rho)-J^{(2)}(\rho)=S(\rho_{c|a})-S(\rho_{c|ab})$. Notice
that $S(\rho_{c|a})=\min_{\{E^a\otimes
I^b\}}S(\rho_{c|\{E^a_l\otimes I^b\}})$, and that $\{E^a_l\otimes
I^b\}$ may not be the optimal POVM $\{E^{ab}_l\}$ in the definition
of $S(\rho_{c|ab})$. Therefore, we have $J^{(3)}(\rho)\geq
J^{(2)}(\rho)$. This completes the proof.

For quantum correlations, there are no fixed quantitative relation
between genuine and pairwise quantum correlations, which we will
illustrate in the next section by some concrete examples.

\section{ANALYTIC EXPRESSION OF GENUINE TRIPARTITE QUANTUM DISCORD FOR RANK-TWO SYMMETRICAL STATES}

We now consider genuine quantum discord of rank-two symmetrical
states of three qubits, which we can get the analytic results. A
rank-two symmetrical tripartite system can be written as
\begin{equation}
\rho=p|\varphi_1\rangle\langle\varphi_1|+(1-p)|\varphi_2\rangle\langle\varphi_2|\label{eq5},
\end{equation}
where $|\varphi_i\rangle$ is a three-qubit symmetrical state. The
state as in Eq. (\ref{eq5}) can be purified to a four-qubit pure
state by attaching an auxiliary system $d$:
\begin{equation}
|\Psi_{abcd}\rangle=\sqrt{p}|\varphi_1,0\rangle+\sqrt{1-p}|\varphi_2,1\rangle.
\end{equation}
Then the Koashi-Winter relation \cite{Koashi:2004} gives
\begin{equation}
D^{(3)}(\rho)=S(\rho_{ab})+E(\rho_{cd})-S(\rho).\label{d3}
\end{equation}
Here $E(\rho_{cd})$ is the entanglement of formation (EOF) between
qubits $c$ and $d$, which is defined as
\begin{equation}
E(\rho_{cd})=\min_{\{p_i,|\phi_i\rangle_{cd}\}}\sum_ip_iS(Tr_c(|\phi_i\rangle\langle\phi_i|)),\label{EOF}
\end{equation}
and can be calculated as follows.
$E(\rho_{cd})=-h\log_2h-(1-h)\log_2(1-h)$, where
$h=\frac{1+\sqrt{1-C_{cd}^2}}{2}$, $C_{cd}$ being the concurrence of
$\rho_{cd}$ \cite{Wootters:1998,Scott:1997}. Optimal POVM
$\{E_i^{ab}\}$ in the definition of $D^{(3)}$ related to the optimal
pure state decomposition $\{p_i,|\phi_i\rangle_{cd}\}$ for EOF as
follows \cite{Koashi:2004}:
\begin{equation}
\rho^i_{c|ab}=Tr_{ab}\{E_i^{ab}\rho\}/p_i=Tr_{d}\{|\phi_i\rangle\langle\phi_i|\}.\label{eq3}
\end{equation}

\begin{figure}[h]
\centering

\includegraphics[width=0.45\textwidth]{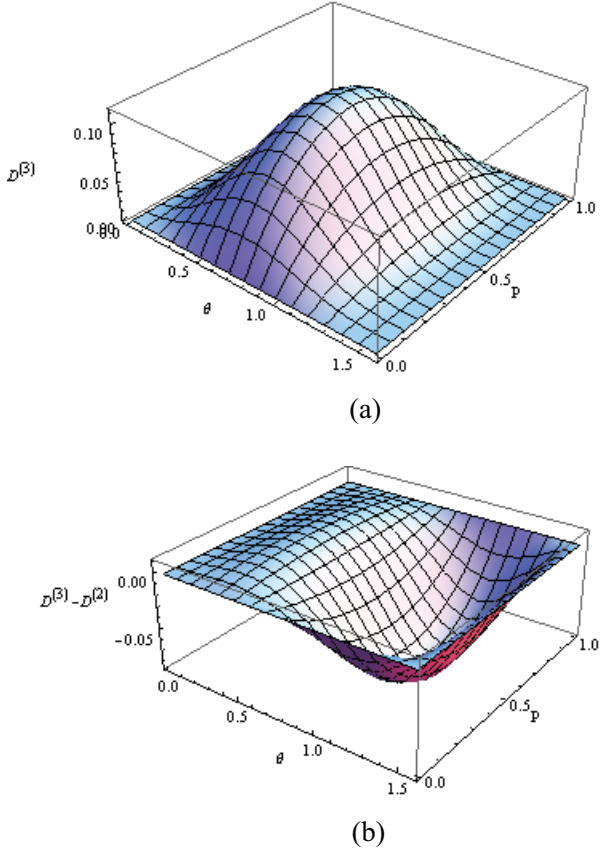}
\caption{(a):$D^{(3)}(\rho)$ as a function of $\theta$ and $p$.
(b):$D^{(3)}(\rho)-D^{(2)}(\rho)$ as function of $\theta$ and $p$.}
\end{figure}

We study two concrete examples to investigate more closely the
properties of genuine tripartite correlations. Firstly, consider a
symmetric tripartite system as the form:
\begin{equation}
\rho=p|000\rangle\langle000|+(1-p)|\phi\phi\phi\rangle\langle\phi\phi\phi|\label{eq7}
\end{equation}
where $|\phi\rangle=\cos\theta|0\rangle+\sin\theta|1\rangle$. This
is a three-qubit product state with no entanglement of any type. We
calculate nonzero eigenvalues of $\rho$ and $\rho_{ab}$ as well as
the analytical formula of $h$
\begin{align}
\lambda_1^{abc}=&\frac{1}{8} (4-\sqrt{2} \sqrt{8-pq(22-15 \cos2\theta-6\cos4\theta-\cos6\theta}),\nonumber\\
\lambda_2^{abc}=&\frac{1}{8} (4+\sqrt{2}\sqrt{8-pq(22-15\cos2\theta-6\cos4\theta-\cos6\theta}),\nonumber\\
\lambda_1^{ab}=&\frac{1}{4}(2-\sqrt{2} \sqrt{2-pq(5-4\cos2\theta-\cos4\theta }),\nonumber\\
\lambda_2^{ab}=&\frac{1}{4}(2+\sqrt{2} \sqrt{2-pq(5-4\cos2\theta-\cos4\theta }),\nonumber\\
h=&\frac{1}{8}(4+(16+15pq(\cos2\theta-18+2\cos4\theta +\cos6\theta )\nonumber\\
&+8\sqrt{pq (\sin\theta)^6\sqrt{pq(5 \sin\theta+\sin3\theta)^2}})^{\frac{1}{2}}),
\end{align}
where $q=1-p$. Then from Eq. (\ref{d3}), we obtain the the analytic
expression for the genuine tripartite quantum discord:
\begin{eqnarray}
D^{(3)}(\rho)&=&-\lambda_1^{ab}\log_2\lambda_1^{ab}-\lambda_2^{ab}\log_2\lambda_2^{ab}
-(1-h)\log_2(1-h)\nonumber\\
&&-h\log_2h+\lambda_1^{abc}\log_2\lambda_1^{abc}+\lambda_2^{abc}\log_2\lambda_2^{abc} \label{eq8}.
\end{eqnarray}

The analytical results are plotted in Fig. 1. Figure $(1a)$ is
$D^{(3)}(\rho)$ as a function of $p$ and $\theta$. We see that the
$D^{(3)}(\rho)$ equals to zero when $p=0$ or $p=1$ and it takes the
maximal value for $p=\frac{1}{2}$, when $\theta$ is fixed. It is not
difficult to find that $D^{(3)}(\rho)$ is symmetric for $p$ with
$p=\frac{1}{2}$ the symmetric axe. This can be understood as
follows. The state in Eq. (\ref{eq7}) can be transformed into
$\rho^\prime
=(1-p)|000\rangle\langle000|+p|\phi\phi\phi\rangle\langle\phi\phi\phi|$
by the unitary operator $U^\prime=U\otimes U \otimes U$, where
$U=\{\cos\theta,\sin\theta;\sin\theta,-\cos\theta\}$. Genuine
correlations are preserved under local unitary operations, that is
$D^{(3)}(\rho)=D^{(3)}(\rho^\prime )$. Hence, $D^{(3)}(\rho)$ is
invariant when $p$ and $1-p$ are interchanged. Then, we find that
$D^{(3)}(\rho)$ takes the maximal value when $p=\frac{1}{2}$ and
$\theta=0.688$. In Fig. 1(b) is plotted the difference between
$D^{(3)}$ and $D^{(2)}$ as a function of $p$ and $\theta$. The cases
which $D^{(3)}(\rho)$ is less than, equal to or greater than
$D^{(2)}(\rho)$ are all possible.

We then turn to find out the corresponding optimal measurements by
which we get the genuine tripartite quantum discord. The case for
$p=\frac{1}{2}$ is discussed in here. Firstly, we achieve the optimal
pure state decomposition of $\rho_{cd}$ which minimized the EOF of
the state using the method in Ref. \cite{Scott:1997}:
\begin{eqnarray}
|\phi_1\rangle&=&\frac{1}{2}(\sqrt{1+{\cos}^2\theta }+{\sin}\theta )|00\rangle\nonumber\\
&&+\frac{1}{2} {\cos}\theta (\sqrt{1+{\cos}^2\theta }-{\sin}\theta )|01\rangle \nonumber\\
&&+\frac{1}{2} (\sqrt{1+{\cos}\theta ^2}-{\sin}\theta ) {\sin}\theta|11\rangle,\nonumber\\
|\phi_2\rangle&=&\frac{1}{2} (\sqrt{1+{\cos}^2\theta }-{\sin}\theta )|00\rangle\nonumber\\
&&+\frac{1}{2}({\cos}\theta \sqrt{1+{\cos}^2\theta }+{\cos}\theta  {\sin}\theta )|01\rangle\nonumber\\
&&+\frac{1}{2} (\sqrt{1+{\cos}^2\theta } {\sin}\theta +{\sin}^2\theta
)|11\rangle.
\end{eqnarray}
Then from Eq. (\ref{eq3}), we have the optimal measurement bases for
$D^{(3)}$ as
\begin{align}
|E^{ab}_1\rangle=&\frac{\sqrt{3+\cos2\theta}+\sqrt2\sin\theta}{2\sqrt2}|00\rangle\nonumber\\
&-\frac{{\sin}\theta(\sqrt{3+\cos2\theta}-\sqrt2{\sin}\theta )}{2\sqrt{3+\cos2\theta}}|11\rangle\nonumber\\
&-\frac{{\cos}\theta (\sqrt{3+\cos2\theta}-\sqrt2{\sin}\theta )}{2
\sqrt{3+\cos2\theta}}(|01\rangle+|10\rangle),
\end{align}
\begin{align}
|E^{ab}_2\rangle=&\frac{\sqrt{3+\cos2\theta}-\sqrt2\sin\theta}{2\sqrt2}|00\rangle\nonumber\\
&+\frac{{\sin}\theta(\sqrt{3+\cos2\theta}+\sqrt2{\sin}\theta )}{2\sqrt{3+\cos2\theta}}|11\rangle\nonumber\\
&+\frac{{\cos}\theta (\sqrt{3+\cos2\theta}+\sqrt2{\sin}\theta )}{2\sqrt{3+\cos2\theta}}(|01\rangle+|10\rangle).
\end{align}

\begin{figure}[h!]
\centering

\includegraphics[width=0.45\textwidth]{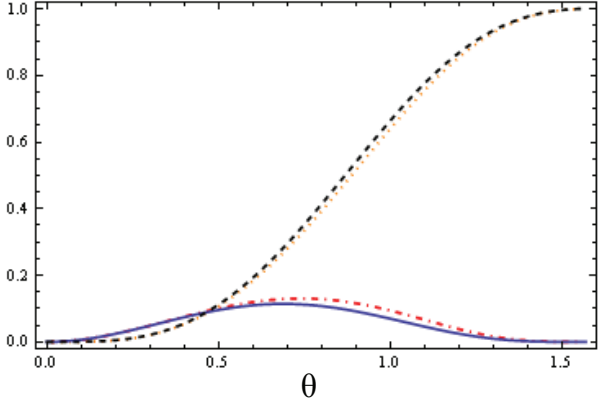}
\caption{Correlations for the state, $\rho=p|000\rangle\langle000|+(1-p)|\phi\phi\phi\rangle\langle\phi\phi\phi|$, vary with $\theta$ when $p=1/2$ : $D^{(3)}(\rho)$(blue solid line),$J^{(3)}(\rho)$ (black dashed line), $D'^{(3)}(\rho)$  (red dot-dashed line), $J'^{(3)}(\rho)$ (orange dotted line).}
\end{figure}
The other two measurements with above two measurements constitute
a set of orthogonal basis which satisfy that
$\sum_{k=1}^4|E^{ab}_k\rangle \langle E^{ab}_k|=I$.  It must be
noticed  $|E^{ab}_k\rangle$ can not always be written as the form
that $|E^{a}_l\rangle \otimes|E^{b}_m\rangle$.

The genuine tripartite classical and quantum correlations we defined
are different with which defined in \cite{Giorgi:2011}. Here we
compare our measure for genuine correlations with those defined in
Ref \cite{Giorgi:2011}, where
$J'^{(3)}(\rho)=S(\rho_{c})-S'(\rho_{c|ab})$ and
$D'^{(3)}(\rho)=S(\rho_{ab})+S'(\rho_{c|ab})-S(\rho_{abc})$ with
$S'(\rho_{c|ab})=\min_{\{E^a_l,E^b_m\}}[S(\rho_{c|\{E^a_l,E^b_m\}})]$
are the genuine quantum and classical correlations. Since
$\{E^a_l\otimes E^b_m\}$ may not be the optimal POVM $\{E^{ab}_m\}$
in the definition of $D^{(3)}$, we have $D^{(3)}(\rho)\leq
D'^{(3)}(\rho)$ and $J^{(3)}(\rho)\geq J'^{(3)}(\rho)$. The
comparison is shown in figure 2 which is the $\theta$-dependent
correlations variation curves when $p=\frac{1}{2}$. From figure 2,
we can see that $D^{(3)}(\rho)$ and $ D'^{(3)}(\rho)$ are quite
close to each other. However, the two measures of genuine quantum
correlation do not coincide for $\theta\neq0$ or $\pi/2$. It means
that even for separable state in Eq. (\ref{eq7}), the optimal
measurement $\{E^{ab}_m\}$ in the definition of genuine quantum
correlation $D^{(3)}$ can not be written as $E^{ab}_m=E^a_l\otimes
E^b_m$. Another interesting phenomenon is that the genuine quantum
correlation $D^{(3)}(\rho)$ may surpass the genuine classical
correlation $J^{(3)}(\rho)$ even for separable states.

\begin{figure}[h]
\centering

\includegraphics[width=0.44\textwidth]{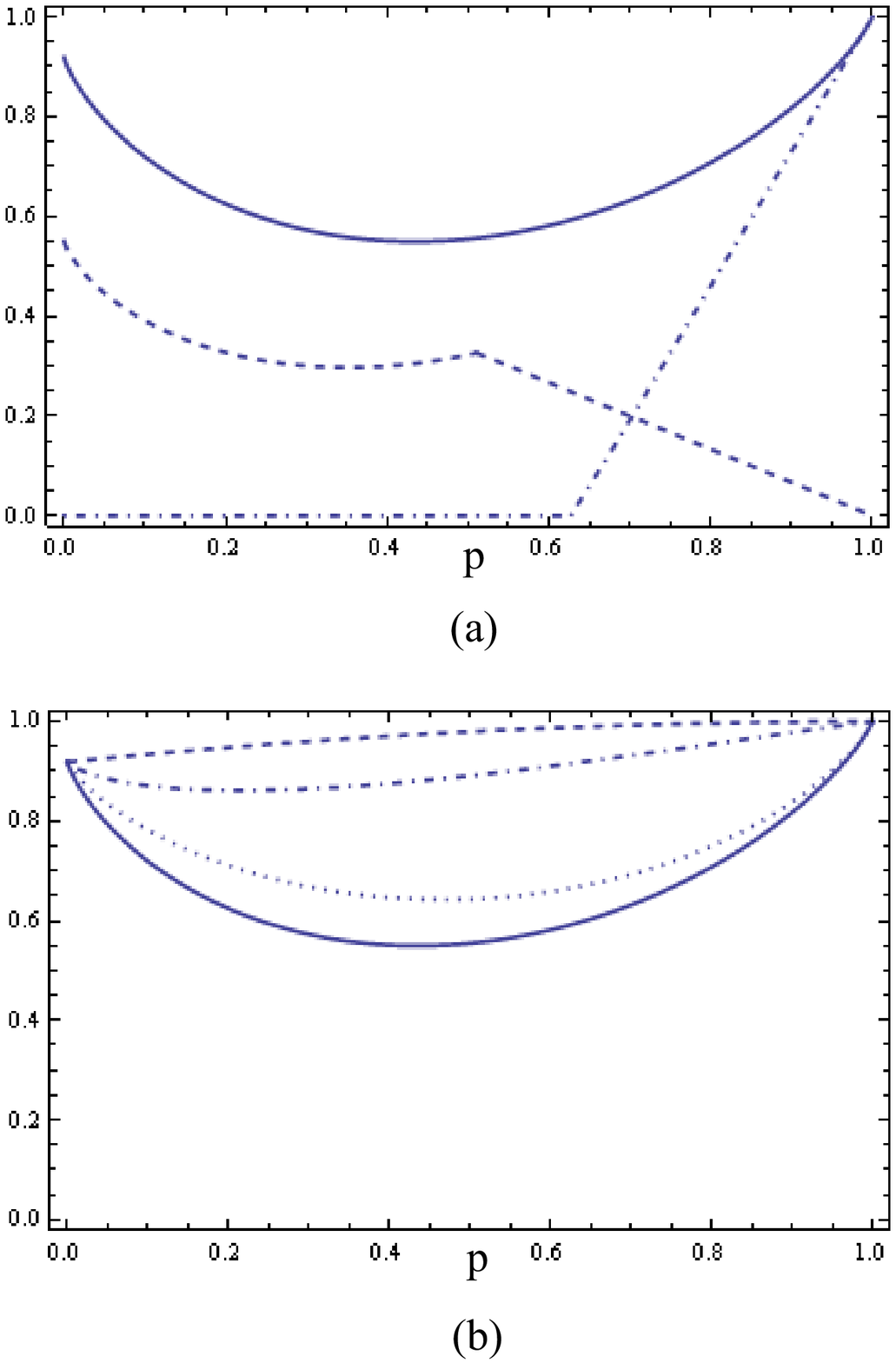}
\caption{Correlations for the state, $\rho_{abc}=p|GHZ\rangle\langle
GHZ|+(1-p)|W\rangle\langle W|$, vary with p. (a): $D^{(3)}(\rho)$
(solid line), $D^{(2)}(\rho)$ (dashed line), $\tau_3$(dot-dashed
line). (b):$D^{(3)}(\rho)$(solid line),$J^{(3)}(\rho)$ (dashed
line), $D'^{(3)}(\rho)$  (dot-dashed line), $J'^{(3)}(\rho)$ (dotted
line).}
\end{figure}

Now, we consider a state of this form
\begin{equation}
\rho=p|GHZ\rangle\langle GHZ|+(1-p)|W\rangle\langle W|\label{eq2},
\end{equation}
where $|GHZ\rangle=\frac{1}{\sqrt{2}}(|000\rangle+|111\rangle)$ and
$|W\rangle=\frac{1}{\sqrt{3}}(|100\rangle+|010\rangle+|001\rangle)$.
The genuine tripartite discord of the state as in equation
(\ref{eq2}) can be calculated using the above method
\begin{eqnarray}
D^{(3)}(\rho)&=&-(\frac{1-p}{3}+\frac{p}{2})\log_2(\frac{1-p}{3}+\frac{p}{2})\nonumber\\
&&-(\frac{2}{3}(1-p))\log_2(\frac{2}{3}(1-p))-(\frac{p}{2})\log_2(\frac{p}{2})\nonumber\\
&&+p\log_2p+(1-p)\log_2(1-p).
\end{eqnarray}
Figure 3(a) shows  $D^{(3)}(\rho)$ is greater than $D^{(2)}(\rho)$.
There is a transition point at $p=0.51$ for $D^{(2)}(\rho)$ because
of the sudden change of the measurement basis. The three-tangle
$\tau_3$ \cite{Coffman:2000} of the state (\ref{eq2}) which we
showed in the Fig. 3(a) has been obtained in ref \cite{Robert:2006}.
We can see that the genuine quantum discord is no less than the
three-tangle $\tau_3$. Fig. 3(b) shows $D'^{(3)}(\rho)$,
$D^{(3)}(\rho)$, $J^{(3)}(\rho)$ and $J'^{(3)}(\rho)$ as functions
of $p$. We can see that the four quantities coincide only for $p=0$
or $p=1$, where the state in Eq. (\ref{eq2}) is just the $|W\rangle$
state or $|GHZ\rangle$ state. For $p\in(0,1)$, we can see that in
this state, the gap between $D'^{(3)}(\rho)$ and $D^{(3)}(\rho)$, as
well as that between $J^{(3)}(\rho)$ and $J'^{(3)}(\rho)$ can be
very large. Moreover, the ordering is different, i.e., $J^{(3)}$ is
greater than $D^{(3)}$ while $J'^{(3)}$ is less than $D'^{(3)}$.

\section{CONCLUSION }
In summary, we have investigated the genuine correlations
in a tripartite quantum state. We propose the definitions for
genuine tripartite quantum and classical correlations, as well as
the analytical expression of them for rank-two symmetrical states of
three qubits. We have shown that, genuine tripartite classical
correlations and quantum discord are both equal to half of genuine
tripartite total correlations in pure tripartite states, which
coincide with the definition of genuine correlation given in
\cite{Giorgi:2011}. For a symmetrical tripartite state, the
quantitative relation between genuine tripartite quantum discord and
its pairwise counterpart is not fixed, while the genuine tripartite
total and classical correlations are no less than their any pairwise
counterparts. Interestingly, the genuine quantum correlation can
surpass the genuine classical correlation even in some separable
states.

The study of various correlations in multipartite states is of interests
not only for quantum information science but also for many-body systems in
condensed matter physics and statistical mechanics. However, no consensual
measures of various correlations in multipartite case are found,
even in the well-studied entanglement case. The correlation measures in
tripartite states proposed in this paper should be a start point in
completely quantifying the multipartite correlations.
It will also be interesting
to use the correlation measures presented in this paper
in some real physical systems.

\emph{Acknowledgements:} This work is supported by ``973'' program (2010CB922904)
and NSFC (10974247,11175248).


\begin{thebibliography}{99}
\bibitem{entanglementreview}R. Horodecki, P. Horodecki, M. Horodecki and
K. Horodecki, Rev. Mod. Phys. {\bf 81}, 865 (2009).

\bibitem{cuinaturecomm}J. Cui, M. Gu, L. C. Kwek, M. F. Santos, Heng Fan and V. Vedral, Nature Commun. {\bf 3}, 812 (2012).

\bibitem{Ollivier:2001}H. Ollivier and W. H. Zurek,  Phys.\ Rev.\ Lett.\ {\bf 88}, 017901 (2001).

\bibitem{vedral1}L. Henderson and V. Vedral, J. Phys. A \textbf{34}, 6899 (2001).

\bibitem{vedral2}V. Vedral, Phys. Rev. Lett. \textbf{90}, 050401 (2003).

\bibitem{Yu:2004}T. Yu and J. H. Eberly, Phys.\ Rev.\ Lett.\ {\bf 93}, 140404 (2004).

 \bibitem{Werlang:2009}T.~Werlang, S. S. Souza, F. F. Fanchini, and C. J. Villas Boas, Phys.\ Rev.\ A {\bf 80}, 024103 (2009).

\bibitem{Datta:2008}A. Datta, A. Shaji, and C. M. Caves, Phys.\ Rev.\ Lett.\ {\bf 100}, 050502 (2008).

\bibitem{Roa:2011}Luis Roa, J. C. Retamal, and M. Ali-Vaccarezza, Phys.\ Rev.\ Lett.\ {\bf 107}, 080401 (2011).

\bibitem{Bo:2012}Bo Li, Shao-Ming Fei, Zhi-Xi Wang, and Heng Fan, Phys.\ Rev.\  {\bf 85}, 022328 (2012).

\bibitem{Ye:2010}Ye Yeo, Jun-Hong An, and C. H. Oh, Phys.\ Rev.\ {\bf 82}, 032340 (2010).

 \bibitem{luo:2008}Shunlong~Luo, Phys.\ Rev.\ A {\bf 77}, 042303 (2008).

\bibitem{Ali:2010}M. Ali, A. R. P. Rau, and G. Aliber, Phys.\ Rev.\ A {\bf 81}, 042105 (2010).


\bibitem{Koashi:2004}M. Koashi, Phys.\ Rev.\ A {\bf 69}, 022309 (2004).

\bibitem{Rulli:2011}C. C. Rulli and M. S. Sarandy, Phys.\ Rev.\ A {\bf 84}, 042109 (2011).

 \bibitem{Agrawal:2011}L. Chakrabarty, P. Agrawal and A. K. Pati, arXiv: 1006,5784. (2011)

  \bibitem{Giorgi:2011}G. L. Giorgi, B. Bellomo, F. Galve, and R. Zambrini, Phys.\ Rev.\ Lett.\ {\bf 107}, 190501 (2011).

   \bibitem{Modi:2010}K. Modi, T. Paterek, W. Son, V. Vedral, and M. Williamson, Phys.\ Rev.\ Lett.\ {\bf 104}, 080501 (2010).

  \bibitem{Bennett:2011}C. H. Bennett, A. Grudka, M. Horodecki, P. Horodecki, and R. Horodecki, Phys.\ Rev.\ A {\bf 83}, 012312 (2011).

\bibitem{Nielsen:2000}M. A. Nielson and I. L. Chuang, Quantum Computation and Quantum Information, (2000)


 \bibitem{Bennett:1996}C. H. Bennett, D. P. Divincenzo, J. A. Smolin, and W. K. Wootters, Phys.\ Rev.\ A {\bf 54}, 3824 (1996).

\bibitem{Wootters:1998}W. K. Wootters, Phys.\ Rev.\ Lett.\ {\bf 80}, 2245 (1998).

\bibitem{Scott:1997}S. Hill and W. K. Wootters, Phys.\ Rev.\ Lett.\ {\bf 78}, 5022 (1997).

\bibitem{Robert:2006}R. Lohmayer, A. Osterloh, J. Siewert, and A. Uhlmann, Phys.\ Rev.\ Lett.\ {\bf 97}, 260502 (2006).

 \bibitem{Coffman:2000}V. Coffman, J. kundu, and W. K. Wootters, Phys.\ Rev.\ A {\bf 61}, 052306 (2000).

















\end{thebibliography}
\end{document}